\documentclass{elsart}

\usepackage{graphicx}

\usepackage{amssymb}

\begin{document}

\begin{frontmatter}

\title{Characterization of spatiotemporal chaos in an inhomogeneous
active medium}

\author[CAB]{S. Bouzat,}
\author[CAB,PMI]{H.S Wio \thanksref{newad},}
\author[BSAS]{G.B. Mindlin}

\address[CAB]{Grupo de F\'{\i}sica Estad\'{\i}stica, Centro
At\'omico Bariloche (CNEA) and Instituto Balseiro (CNEA and UNC),
8400-San Carlos de Bariloche, Argentina}
\address[PMI]{Departament de F\'{\i}sica, Universitat de les
Illes Balears and IMEDEA (CSIC-UIB),  E-07122 Palma de Mallorca,
Spain}
\address[BSAS]{Departamento de F\'{\i}sica, FCEN, UBA, Ciudad
Universitaria, Pab. I,  C.P.(1428), Buenos Aires, Argentina}
\thanks[newad]{New Address: Instituto de
F\'{\i}sica de Cantabria (UC-CSIC), Avda Los Castros s/n, 39005
Santander, Spain, E-mail: wio@ifca.unican.es}


\begin{abstract}
We study a reaction diffusion system of the activator-inhibitor
type with inhomogeneous reaction terms showing spatiotemporal
chaos. We analyze the topological properties of the unstable
periodic orbits in the slow chaotic dynamics appearing, which can
be embedded in three dimensions. We perform a bi-orthogonal
decomposition analyzing the minimum number of modes necessary to
find the same organization of unstable orbits.
\end{abstract}

\begin{keyword}
Active Media \sep Spatiotemporal Chaos \sep Bi-Orthogonal
Decomposition

05.45.-a \sep 47.54.+r \sep 47.52.+j
\end{keyword}

\end{frontmatter}

\section{\bf Introduction}

Spatiotemporal chaos \cite{general1} has been extensively studied
within the context of coupled maps, the complex Ginzburg--Landau
equation, the Kuramoto--Shivashinsky equation and other related
equations \cite{chaos1}. However, studies of spatiotemporal chaos
in reaction--diffusion models closely connected to experimental
systems are scarce. Here we analyze the characteristics of the
chaotic dynamics recently found in a simple inhomogeneous
reaction--diffusion system \cite{PLAinh} of the the type used to
describe chemical reactions in gels \cite{gels} and patterns in
coupled electrical circuits \cite{purw}.

Among the main issues in the study of spatiotemporal chaos we can
select those related to clarifying some aspects of the relation
between pattern formation and chaos as well as the low dimensional
description of the chaotic behavior. The latter aspect, that is
understanding that physically continuous systems with an infinity
of degrees of freedom (spatially extended systems) usually show
temporal behavior that can be well described by models with few
degrees of freedom, is of extreme relevance. In this context there
arise some questions. For instance, in low dimensional dynamical
systems, chaotic solutions coexist with unstable periodic orbits
which constitute the backbone of the strange attractor: could some
orbits be extracted from the time series of our extended system?
and, is the complex time evolution of the system of a
dimensionality small enough to be understood in terms of simple
stretching and folding mechanisms?

In order to investigate these questions within reaction--diffusion
systems, we have analyzed the same simple, inhomogeneous,
activator-inhibitor model discussed in Ref. \cite{PLAinh}. It is
worth remembering here that reaction--diffusion models of the
activator--inhibitor type have provided a useful theoretical
framework for describing pattern formation phenomena with
applications ranging from physics to chemistry, biology and
technology \cite{murray,mik,general2}.

In Ref. \cite{PLAinh} by introducing spatial dependence of the
parameters of the activator--inhibitor equations, a system in
which different parts of the media do not share the same reaction
properties was modelled. The case considered corresponds to a
finite one dimensional oscillatory medium with an immersed
bistable spot. In that system, in addition to stationary,
Hopf--like and Turing--like patterns, quasiperiodic inhomogeneous
oscillations and spatiotemporal chaos were also found. In Ref.
\cite{PREinh}, different generalizations of the system
(bidimensional versions) have been studied. In the present paper
we analyze the dynamics of the same one dimensional system in the
quasiperiodic and chaotic regions. More specifically, one of our
main aims is to understand the topological properties of the
chaotic dynamics. The model is given by the reaction--diffusion
equations
\begin{eqnarray}
\label{sys}
\dot{u}&=&\partial^2_x u-u^3+u-v            \nonumber \\
\dot{v}&=&D_v \partial^2_x v + u -\gamma v,
\end{eqnarray}
which describe a bistable medium for $\gamma>1$ and an oscillatory
one for $\gamma<1$. In order to model the inhomogeneous situation
of a bistable domain immersed in an oscillatory medium, a spatial
dependence of this parameter is introduced setting
$\gamma=\gamma(x)\equiv .9+5\,\exp(-10 \, x^4)$ \cite{coment1}.
This leads to $\gamma \simeq .9 < 1$  for $|x|>.8$ (oscillatory
medium) and $\gamma>1$ for $|x|<.8$ (bistable medium). As was done
in \cite{PLAinh}, we here consider a finite one--dimensional
domain ($-L \le x \le L$) with non--flux boundary conditions in
$\pm L$ and homogeneous initial conditions belonging to the
homogeneous limit cycle that exists for the case $\gamma=.9$. This
choice of the initial state corresponds to the description of an
initially homogeneous oscillatory medium whose reaction properties
are suddenly modified in a localized region.

In the central bistable region the fields converge rapidly  to
values close to those corresponding to one of the two natural
states of the bistable medium ($u_{\pm}\simeq\pm .8,
v_{\pm}\simeq\pm .14$) (chosen depending on the initial
condition), and continue performing small amplitude oscillations
around those values. Hence, there is a spontaneous symmetry
breaking which is ``inherited'' from the properties of the
(uncoupled) bistable medium. Note that the equations of the model
are symmetric under the simultaneous changes $u \to -u$, $v \to
-v$. The rest of the system evolves to different asymptotic
behaviors depending on the parameters $L$ and $D_v$ as indicated
in Figure 1, and described in detail in \cite{PLAinh}.

\begin{figure}
\centering
\resizebox{.8\columnwidth}{!}{\includegraphics{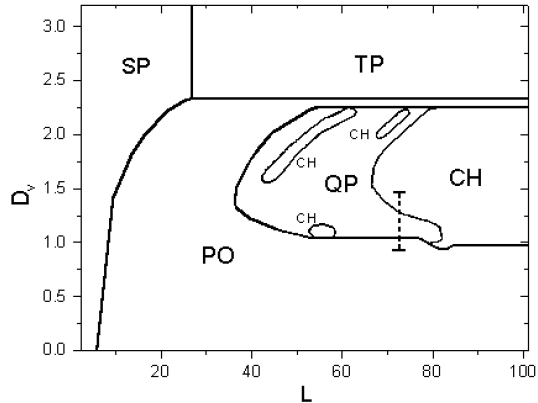}}
\caption{Phase diagram indicating the different asymptotic regimes
in the $(L,D_v)$ plane: stationary patterns (SP), Turing patterns
(TP), periodic inhomogeneous oscillations (PO), quasiperiodic
oscillations (QP) and spatiotemporal chaos (CH). The vertical
dotted segment indicates the region of main interest for this
work.} \label{fig1}
\end{figure}

All the numerical calculations have been done as follows. First,
the system of partial differential equations was approximated by a
system of coupled ordinary differential equations, obtained by a
finite difference scheme. Then the resulting equations were solved
by a Runge--Kutta method of order $2$. Different space and time
discretization schemes were employed in order to check the
results.

The organization of the paper is the following. In the next
Section we show that, in the quasiperiodic and chaotic regimes,
there are two dynamical time scales, a fast and a slow one. We
show that it is possible to find segments of the time series of
the slow dynamics which approximate unstable periodic orbits and
study the organization of the orbits. In Section III, we present
the biorthogonal decomposition of the spatiotemporal time series,
and show that it is possible to capture the main features of the
chaotic dynamics by considering a small number of modes. In the
last Section we present our conclusions.

\section{Characterization of the slow chaotic dynamics: analysis
of the unstable periodic orbits.}

In the non--stationary regions of the phase diagram shown in
Figure 1, the time evolution of the fields $u$ and $v$ is
classified as periodic, quasiperiodic or chaotic \cite{PLAinh}.
Here, we analyze the transition from periodic oscillations to
chaotic behavior along the line indicated with a vertical dotted
segment in Figure 1. We fix $L=72$, for which the dynamics
corresponds to inhomogeneous periodic oscillations for $D_v<1$,
quasiperiodic oscillations for $1<D_v<1.3$, and spatiotemporal
chaos for $1.3<D_v<2$. (For $D_v>2$ the quasiperiodic and periodic
behaviors appear again and for $D_v>2.3$ stationary Turing
patterns arise.) To begin, we will mainly focus our attention on
the chaotic region.

As a measure of chaoticity, in \cite{PLAinh}, the sensibility to
initial conditions was computed. It is important to notice that
the time series displayed a common feature: a fast oscillation of
the field (at the natural frequency of the oscillatory medium),
eventually modulated by a slow varying amplitude. It is the
dynamics of this amplitude what we will analyze here.
In order to study this slow dynamics we record the times
$t_n$ ($n=1,2,...$) at which the $u$--field at $x=L$ reaches a
local maximum as function of $t$ (i.e. when $ \partial_t
u(x,t)|_{x=L}=0$ and $ \partial^2_t u(x,t)|_{x=L}<0$ holds
simultaneously), and analyze the values of $u$ for these times at
different spatial positions. This is equivalent to taking a
Poincare section, and is a way of averaging the fast time scales.

The difference $t_n-t_{n-1}$ is of the order of the natural
period of the oscillatory medium ($\tau_0=14.6$), but slightly
larger (in general, it fluctuates between $\tau_0$ and $20$) as
the oscillations are slowed down by the presence of the bistable
inhomogeneity. In the periodic region, $t_n-t_{n-1}$ converges to
the period of the motion as $n \to \infty$.

Typical time series are shown in Figure 2. In Figure 2a (c),
the time evolution of $u(L)$ is displayed for a parameter value
at which the system behaves quasiperiodically (chaotically). The
slow varying amplitude is shown in Figure 2b (d), where we have
plotted the values of the maxima of $u(L)$ as a function of $t$
(i.e. $u(L)$ measured at times $t_n$).

\begin{figure}
\centering
\resizebox{.5\columnwidth}{!}{\includegraphics{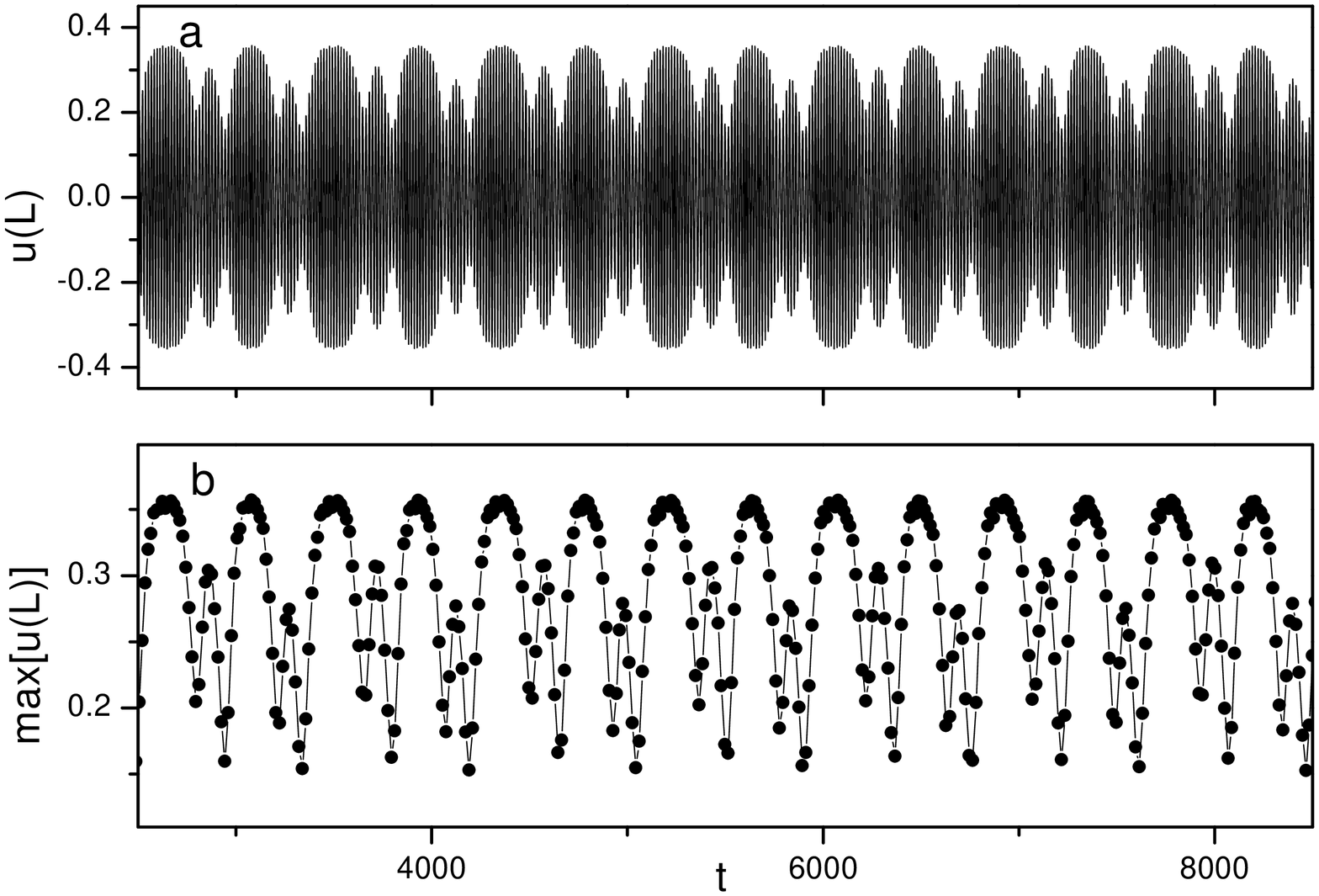}}
\resizebox{.5\columnwidth}{!}{\includegraphics{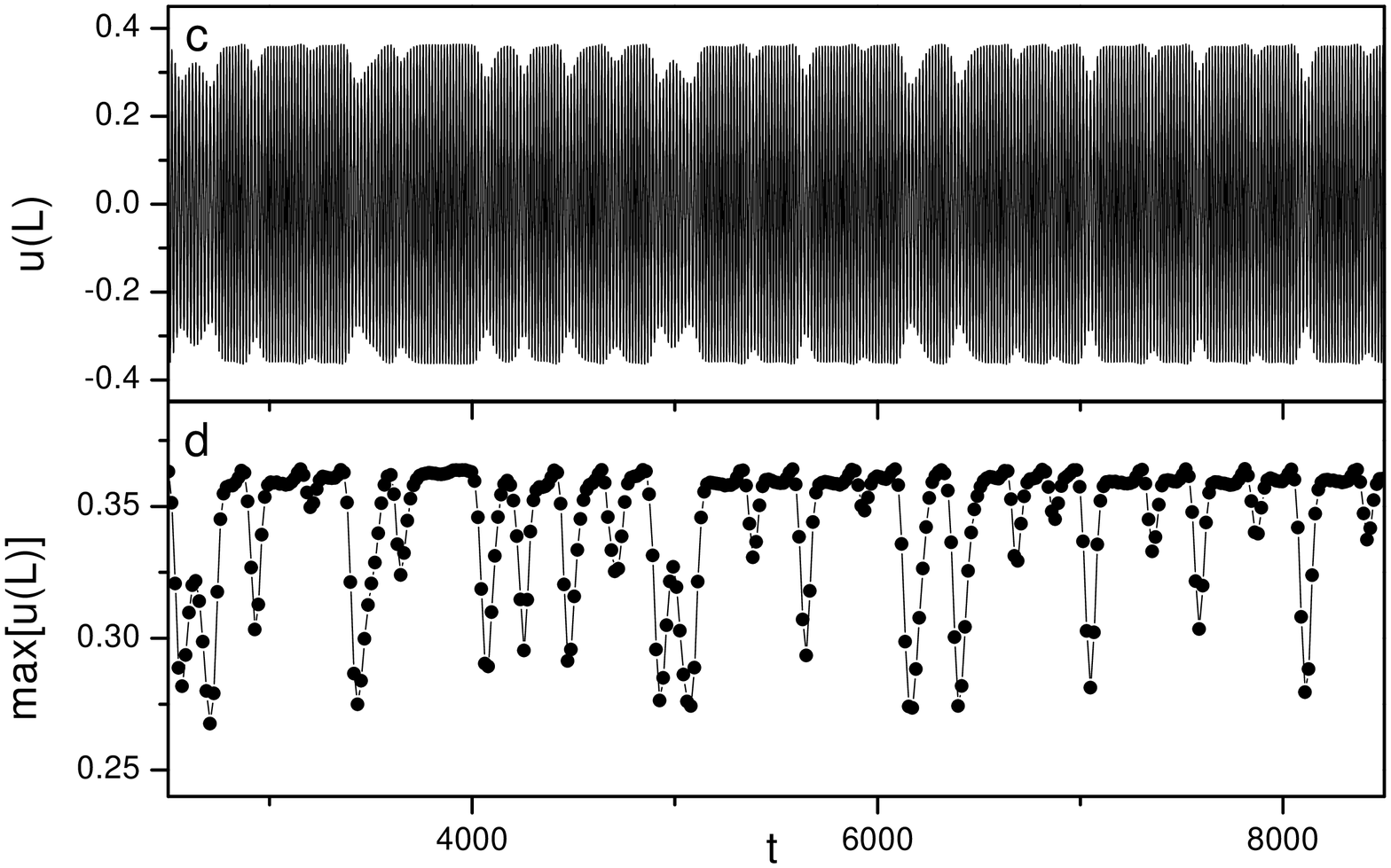}}
\caption{(a) Evolution of $u(L)$ for $D_v=1.1$ (quasi-periodic
regime) in a time window in the asymptotic regime. (b) Slow
dynamics: plot of the maxima of $u(L)$ corresponding to the same
time window. (c) and (d), ibidem figures (a) and (b) for $D_v=1.4$
(chaotic regime).} \label{fig2cd}
\end{figure}

In general, in low dimensional dynamical systems, chaotic
solutions coexist with unstable periodic orbits which constitute
the backbone of the strange attractor. We will see that, in our
system, it is possible to extract approximations of periodic
unstable orbits from the time series of the mentioned slow
dynamics, and that the analysis of the organization of these
orbits shows that the chaotic dynamics is low--dimensional.

We begin by defining as {\em reconstructed periodic orbits} the
segments of the time series which can be used as surrogates of the
unstable periodic orbits of the system. These segments are chosen
if they pass a close return test \cite{mind92}. More precisely, if
$y(i)$ represents the data, a close return is a segment of $p$
points beginning at the $i^{th}$ position of the file, for which
$y(i+k) \approx y(i+k+p)$ for $k=1,2,...$. In this notation, $p$
is called the period.

We have looked for unstable orbits at the whole time series of the
slow dynamics of the $u$ field (data taken at times $t_n$) at four
different positions: $x_0=0$, $x_1=14$ (approximately one Turing
wavelength away the bistable domain), $x\simeq x_2=L/2=36$, and
$x=L$.

In Figure 3a(c) we display a segment of period 2(4) taken from a
time series corresponding to data at $x=L$. An embedding of the
data (a multivariate environment created using time delays) is
shown in Figure 3b(d). In Figure 4a and 4b we show the embedding
of two different reconstructed segments of periods two and three
respectively, coming from data at $x=x_1$. It is worth mentioning
that the unstable periodic orbits do not have properties
corresponding to the inversion symmetry of Eqs. (\ref{sys})
because of the symmetry breaking of the solutions and also because
of the ``stroboscopic'' observation of the dynamics. In Figure 4c,
we show a more complex reconstructed periodic orbit coming also
from data at $x=x_1$. Since we have no elements to conjecture that
the chaotic dynamics can live in three dimensions, it could be
argued that embedding the segments in a three dimensional space
might not be useful. Yet, if the reconstructed shows some kind of
geometrical organization it would be a most valuable indication of
the geometric process taking place in a small dimensional manifold
within the available phase space.

\begin{figure}
\centering
\resizebox{.6\columnwidth}{!}{\includegraphics{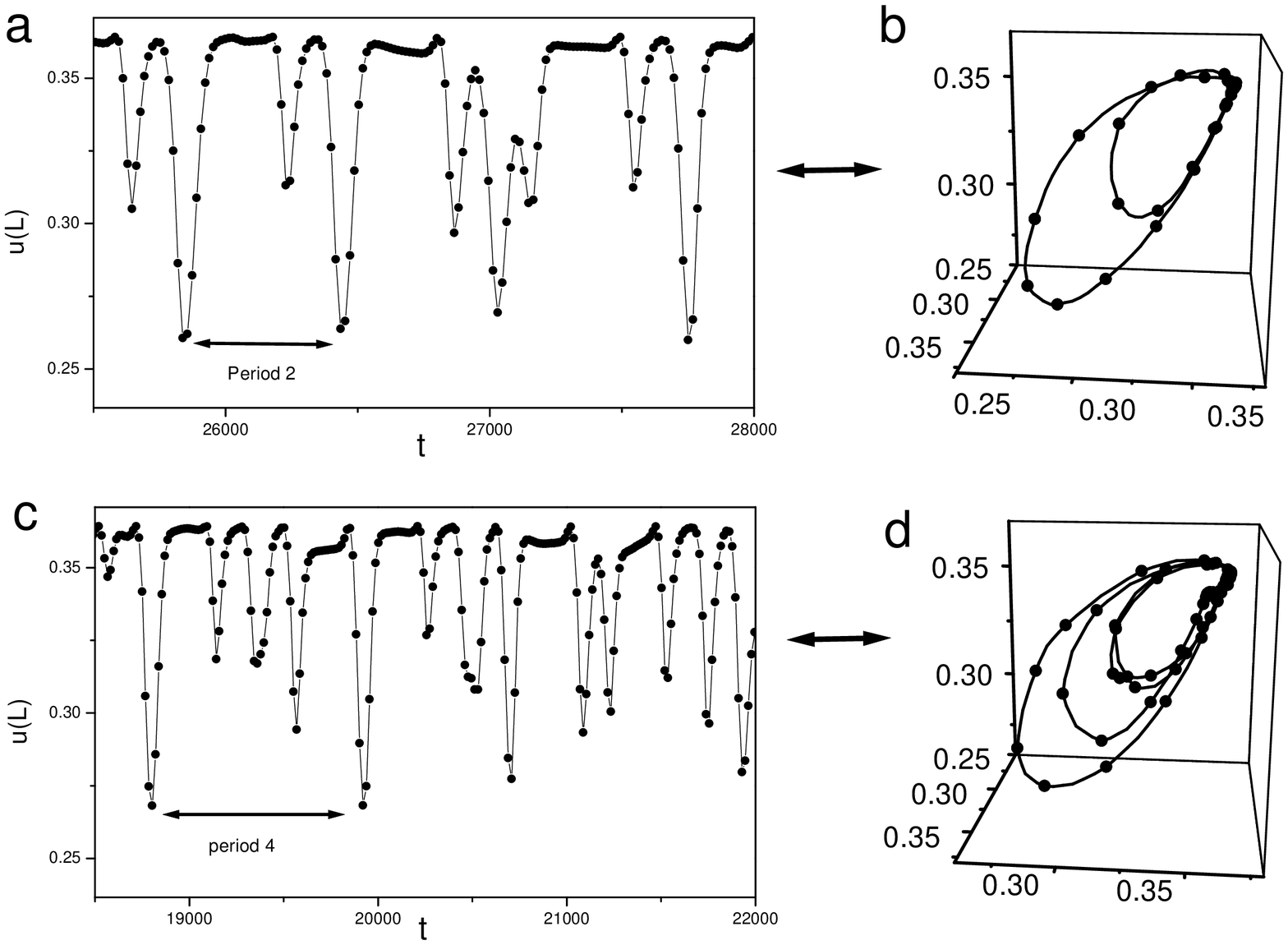}}
\caption{(a) Slow dynamics: values of the field $u(L)$ at times
$t_n$ for a time window in which a segment corresponding to a
period--2 orbit appears. Note that the segment is repeated almost
twice. (b) Period--2 orbit corresponding to the segment indicated
in (a) obtained from an embedding in three dimensions using time
delays (the axis correspond to the data taken at $t_n,t_{n+1}$ and
$t_{n+2}$). (c) Idem (a) for a period--4 orbit. (d) Embedding of
the segment of figure (c). All the data correspond to a simulation
done for $D_v=1.4$.} \label{fig2}
\end{figure}

It is possible to see that the orbits of Figures 3b and 3d wind
around each other as expected if they were related by a period
doubling bifurcation. The topological organization of the orbits
is quantitatively described in terms of their relative rotation
rates and self relative rotation rates. These numbers aim at
describing the way in which the orbits wind around each other
\cite{gilm00}. In order to do so, the curves are given an
orientation, and in a two dimensional projection, a record is made
of which segments pass over which in the original embedded orbits.
In terms of these indices, the relative rotation rates are
computed as explained in \cite{gilm00}. For the period two orbit
of Figure 4a, the self relative rotation rate is
$srrr=-\frac{1}{2},0$, for the period three orbit of Figure 4b, it
is $srrr=(-\frac{1}{3})^2,0$, and the relative rotation rate
between the orbits of period two and three was found to be
$rrr=-\frac{1}{3}$. Notice that this organization is compatible
with a horseshoe mechanism \cite{gilm00}, and that this mechanism
includes the signature of period doubling.


A challenge exists in order to find a simple geometrical
mechanism responsible for the creation of the orbit displayed in
Figure 4c. This orbit can not be placed in a horseshoe template.
Yet, recently, a classification of templates was proposed for
covering the Smale horseshoe \cite{LetGil}. We have observed that
the orbit of Figure 4c can be placed in one of such geometric
objects, which is one of the four inequivalent four-branched {\it
double covers} with rotation symmetry of the Smale. More
specifically, the one identified with topological indices
$(n_0,n_1)=(1,0)$ \cite{LetGil}. This template can also hold any
orbit of the Smale horseshoe template.

\begin{figure}
\centering
\resizebox{.6\columnwidth}{!}{\includegraphics{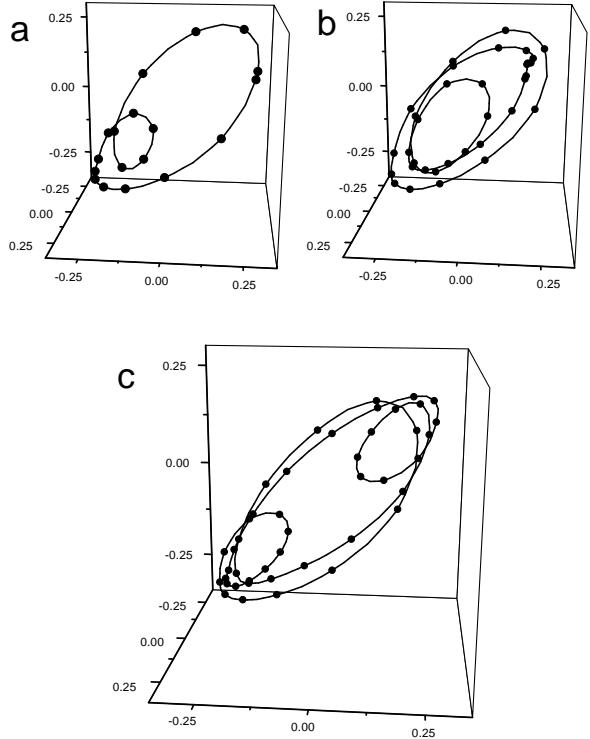}}
\caption{Periodic orbits reconstructed from data taken at $x=x_1
\equiv 14$ at the times $t_n$ using an embedding of time delays
with axis $t_n,t_{n+1},t_{n+2}$. (a) Period--2 orbit. (b)
Period--3 three orbit. (c) Complex periodic orbit.} \label{fig4}
\end{figure}

However, it can not be expected that such template correctly
describes the whole dynamics of the slow varying amplitude. This
is because it is not possible that a rotation symmetry appear when
using a delay embedding. Hence, the embedded attractor must have a
different symmetry or not symmetry at all, and it is expected that
other unstable periodic orbits exist (different to the ones we
have found and with no rotation symmetry).

Note that, when observing the $u$--field at times $t_n$, it is
found that the scales over which it varies are quite different at
the four studied positions ($x_0,x_1,x_2$ and $x_L$): at $x=L$,
$u(t_n)$ oscillates between $.2$ and $.35$ (since we are watching
only the times at which $u(L,t)$ is maximum); at $x=x_1$ and
$x=x_2$, $u(t_n)$ take values in a more or less symmetric way
between $\pm .35$ (in the whole range of the free limit cycle); at
$x=x_0$ (in the bistable domain) the oscillations are of much
smaller amplitude (typically two orders of magnitude), and are not
centered at zero. We remark that, in spite of these differences in
the metrical properties of the dynamics at the several positions,
the organization of the unstable periodic orbits that we have
found is the same everywhere. (In the four positions we find the
same kind of orbits, including the one of Figure 4c.) However,
there are some differences in the frequency of occurrence of the
orbits: note that, we have orbits with the ``small curl'' upward
(as in Figure 3b) or downward (as in Figure 4a). The same two
possibilities appear for orbits of periods three and four. For the
cases of the signal taken at $x_0,x_1$ and $x_2$, the orbits of
period two and three occur preferably with the small curl
downward, while, for $x=L$ they occur (almost always) with the
small curl upward. This is found independently of whether the
fields in the bistable domain converge to negative or positive
values.

The observation that the organization of unstable periodic orbits
is more complex, but some how related to the one of the Smale
horseshoe, gives a hint of what kind of periodic orbits can
eventually be found as parameters are changed. For example, it
suggest that a period doubling sequence may occur in the
transition from the periodic regime to the chaotic regime. With
this in mind, we revisited in detail the transition zone in the
phase diagram of the system going from the periodic region to the
chaotic one along the segment indicated in Figure 1. A period
doubling of the slow dynamics can be clearly identified, as can be
seen in Figure 5.

\begin{figure}
\centering
\resizebox{.6\columnwidth}{!}{\includegraphics{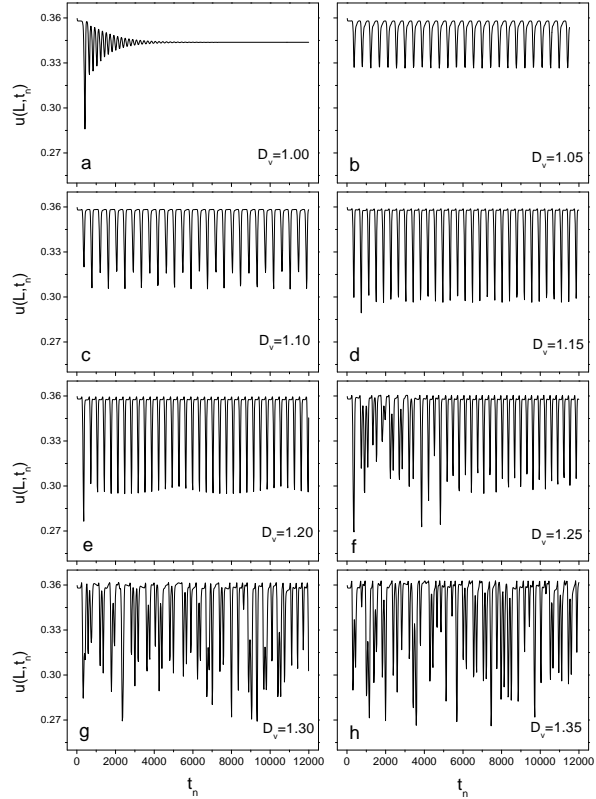}}
\caption{Slow dynamics time series showing a period doubling
sequence. Plots of $u(x=L)$ at times $t_n$ for different values of
$D_v$ (indicated in the figure) following the transition from
periodic regime to quasi-periodic and chaotic. Figure (a)
corresponds to the periodic region where a fixed point occurs in
the slow dynamics. Figures (b) to (e) correspond to the
quasi-periodic region, orbits of period 1,2,4 and 16 are
respectively observed in the slow dynamics. Figure (f) corresponds
to a quasi-periodic regime showing a chaotic-like transient
followed by high-period orbit (not identifiable in the figure).
Figures (g) and (h) correspond to the chaotic regime. Al
calculations are for $L=72$.} \label{fig5}
\end{figure}

The analysis made of the slow dynamics of our extended system
showed that the high dimension of the phase space is not fully
explored. On the contrary, an important collapse of dimensionality
takes place. In the next section we investigate the minimum number
of spatial (linear) modes approximating the spatiotemporal
dynamics that are required in order to recover the topological
organization of unstable periodic orbits observed in the slow
dynamics.

\section{Biorthogonal decomposition.}

It is not easy to know a priori which is the number of spatial
modes which are activated as the dynamics becomes non trivial. In
our problem, we only know that at least three modes should be
active in order to account for the complex behavior described in
the previous section. A method exists to unveil the active
structures in a spatiotemporal problem: the biorthogonal
decomposition (BOD) \cite{bod1,bod2}. This is the most efficient
linear decomposition scheme, in the sense that there is no other
linear decomposition able to capture, with a smaller number of
modes, the same degree of approximation. In our system, the BOD
for the spatiotemporal signal $(u(x,t),v(x,t))$ is given by
\begin{equation}
\label{bod}
(u(x,t),v(x,t))=\sum_{k=1}^{\infty} \alpha_k \psi_k(t) \vec\phi_k(x),
\end{equation}
where the $\alpha^2_k$ (with $\alpha_1 > \alpha_2 >...> 0$) are
the eigenvalues of the temporally--averaged two point correlation
matrix \cite{bod1}, the $\vec\phi_k(x)=(\phi_{u k}(x),\phi_{v
k}(x))$ are the corresponding eigenfunctions (called topos), and
the $\psi_k(t)$ (called chronos) are given by
\begin{equation}
\psi_k(t)=\frac{1}{\alpha_k}\int_{0}^L \left( u(x,t)
\phi_u(x)+v(x,t) \phi_v(x) \right) dx.
\end{equation}

We have observed that, for the system (\ref{sys}), the main
differences in the BOD corresponding to the three dynamical
regimes appear in the chronos and that the spatial modes are
similar in all the three cases. However, we have neither studied
in detail the BOD along the transition to chaos nor analyzed the
question of modes' competition. Our analysis was mainly focused on
finding the number of modes that are necessary to recover the
topological organization of unstable orbits for the chaotic
situation presented in the previous section.

In Figure 6a we show the eigenvalues of the BOD computed for three
different points along the transition line indicated in Figure 1:
a periodic case, a quasiperiodic case and the chaotic situation
studied in the previous section. In Figure 6b we show the first
four topos for the chaotic case. No significant differences are
observed in the spatial modes corresponding to the three regimes.
We have observed that, in all the three regimes, the $i-th$ mode
has $i-1$ spatial nodes (that is, the spatial points where
$u(x)=0$), see Figure 6b. Also, in all the three cases, the second
mode is quasi--stationary and it mainly contributes to the
formation of the fields' profiles in the (quasi--stationary)
bistable region. (Note that the topo 2 in Figure 6b contributes
only around the bistable region ($x\sim 0$)).

\begin{figure}
\centering
\resizebox{.5\columnwidth}{!}{\includegraphics{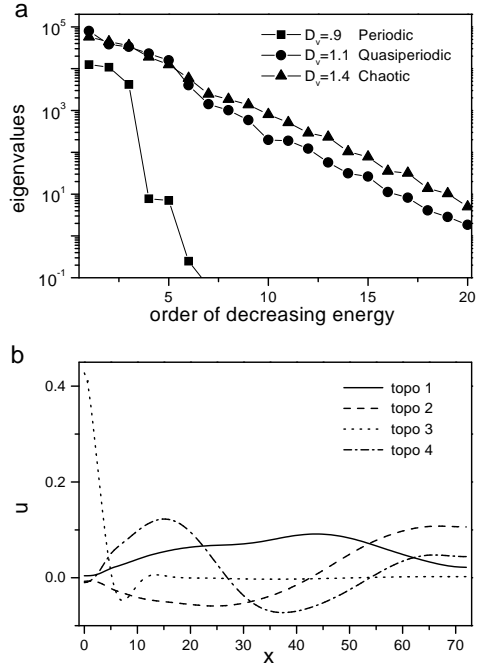}}
\caption{Bi-orthogonal decomposition: (a) Eigenvalues for three
different values of $D_v$ corresponding to chaotic, quasi-periodic
and periodic regimes. (b) $u$-field components of the four
principal topos (corresponding to the four larger eigenvalues) for
the chaotic case with $D_v=1.4$.} \label{fig6}
\end{figure}

In the case of periodic motion, in order to reconstruct the
trivial topology of a single periodic orbit, only the first mode
is necessary (which gives a quasi--homogeneous periodic
oscillation). Moreover, in this region, we have observed that the
whole spatiotemporal dynamics (that is, the periodic wave
propagation phenomenon) can be highly accurately described by
considering an expansion with only three modes (\ref{bod}), as it
is suggested by Figure 6a. Regarding the description of
quasiperiodic and chaotic motion, it requires a higher number of
modes, as can be inferred from Figure 6a. In these regimes all the
chronos seems to be non periodic (excepting the second, which is
constant up to a good approximation).

Finally, for the chaotic case analyzed in the previous section, we
have reconstructed the dynamics of the system using different
numbers of modes. The main result of our analysis is that the
minimum number of modes required to recover the topological
organization of orbits is five. This means that using five modes
(and not four) we were able to recover the orbits presented in the
previous section. There seems to be neither something special in
this number, nor a way of having predicted it a priori. However,
it is important to point out that the extended system under study
can in principle display an infinite dimensional dynamics, and
yet, it dynamically collapses to a five dimensional system which
describe the dynamics of the amplitudes of the linear modes.
Furthermore, it is remarkable that the fact that five modes are
active does not imply that the dimensionality of the observed
strange attractor is larger than four. On the contrary, the
topological organization of the approximated unstable periodic
orbits clearly suggest a lower dimensionality.

\section{Conclusions}

In this work we studied the spatiotemporal solutions of a
reaction-diffusion system of the activator-inhibitor type. Despite
the infinite number of possible degrees of freedom, we have found
that the complex dynamics that emerges can be described in terms
of a small number of modes. The activated modes are coherent
structures which were computed from the simulations of this
extended problem. By separating the dynamics over two time scales,
we observed that the origin of the chaoticity lies on the behavior
of the slow time scale dynamics. The study of these time series
showed not only that the system behaves as a small dimensional
dynamical system, but also suggest that this dynamics may be
understood in terms of simple geometrical process related to the
Smale horseshoe. In fact, a branched manifold recently described
in the literature can hold all the approximated unstable orbits
that we were able to reconstruct.  However, symmetry reasons
indicate that the true mechanism should not be exactly the one
corresponding to that template. The description of the dynamics in
terms of a simple geometric structure not only highlights the
collapse of dimensionality, but it also allowed us to predict the
existence of specific solutions for unexplored regions of
parameter space, such as the reported period doubling sequence.

\vspace{2cm}

{\bf ACKNOWLEDGMENTS:\\} To Ver\'onica Grunfeld for reading the
manuscript.  Partial support from ANPCyT and UBACyT, Argentine
agencies, is also acknowledged. SB and GB thanks for the kind
hospitality extended to them during their stays at the DF (UBA)
and IB-CAB respectively. HSW thanks the MECyD, Spain, for an award
within the {\it Sabbatical Program for Visiting Professors}, and
to the Universitat de les Illes Balears for the kind hospitality
extended to him.


\begin{thebibliography}{99}
\bibitem{coment1} The exact shape of this function is unimportant,
as was discussed in \cite{PLAinh} and \cite{PREinh}.
\end{thebibliography}
\end{document}